\documentclass[prd,aps,nofootinbib,showpacs]{revtex4-2} 

\usepackage{amsmath,amssymb,bm,graphicx}
\usepackage{hyperref}
\usepackage{siunitx}
\usepackage{physics}

\newcommand{\tc}{\tilde{c}}

\newcommand{\ttt}{\tilde{t}}

\newcommand{\tH}{\tilde{H}}

\newcommand{\tD}{\textrm{3D}}

\begin{document}

\title{Geometric Interpretation of the Redshift Evolution of $H_0(z)$}

\author{Seokcheon Lee}
\email{skylee@skku.edu}
\affiliation{Department of Physics, Institute of Basic Science, Sungkyunkwan University, Suwon 16419, Korea}

\date{\today}

\begin{abstract}
Recent analyses of the Master Type~Ia supernova (SN~Ia) sample have revealed a mild redshift dependence in the inferred local Hubble parameter, often expressed as $\tilde{H}_0(z)=H_0(1+z)^{-\alpha}$,  where $\alpha$ quantifies possible departures from the standard cosmological time-dilation relation.  In this work, we show that such an empirical scaling can be interpreted as a purely geometric effect arising from a small, gauge-dependent normalization of cosmic time within the Robertson–Walker metric.  
This interpretation naturally unifies the observed redshift evolution of $\tH_0(z)$ and the corresponding deviation in SN~Ia light-curve durations under a single geometric time-normalization framework.  We demonstrate that this mapping leaves all background distances—linked to the Hubble radius in the general-relativistic frame—unchanged, while the apparent evolution in SN~Ia luminosity distances arises from the redshift dependence of the Chandrasekhar mass.  The result provides a unified and observationally consistent explanation of the mild Hubble-tension trend as a manifestation of the geometric structure of cosmic time rather than a modification of the expansion dynamics.
\end{abstract}

\maketitle

\tableofcontents

\section{Introduction}

The so-called Hubble tension refers to the persistent and statistically significant discrepancy between the value of the Hubble constant $H_0$ inferred from early-Universe probes, such as the cosmic microwave background (CMB), and that obtained from late-Universe distance-ladder measurements~\cite{Perivolaropoulos:2021jda,Freedman:2021ahq,Abdalla:2022yfr,Verde:2023lmm,Kamionkowski:2022pkx,Perivolaropoulos:2024yxv,CosmoVerseNetwork:2025alb}.  Within the $\Lambda$CDM cosmological model, the Planck 2018 results yield $H_0 = 67.4 \pm 0.5~{\rm km\,s^{-1}\,Mpc^{-1}}$~\cite{Planck:2018vyg},  whereas the most recent SH0ES Cepheid-calibrated supernova analysis reports $H_0 = 73.0 \pm 1.0~{\rm km\,s^{-1}\,Mpc^{-1}}$~\cite{Riess:2021jrx,Brout:2022vxf}.  This $\sim5\sigma$ discrepancy has proven remarkably stable across independent datasets and calibration techniques, and remains one of the most pressing challenges in contemporary cosmology~\cite{Freedman:2021ahq,DiValentino:2021izs}.  

Beyond the direct $H_0$ values, the tension is also reflected in multiple observational channels.  Baryon acoustic oscillations (BAO) combined with the CMB sound-horizon prior $r_d$ favor the lower Planck value, while strong-lensing time delays~\cite{H0LiCOW:2019pvv,Birrer:2020tax}, the tip-of-the-red-giant-branch (TRGB) distance scale~\cite{Freedman:2021ahq}, and standardizable-candle methods employing gamma-ray bursts and quasars~\cite{Dainotti:2023pwk,Lusso:2020pdb} tend to prefer intermediate or SH0ES-like results.  Such cross-probe inconsistencies suggest either subtle unaccounted systematics or a deeper, geometric origin for the observed tension~\cite{Kamionkowski:2022pkx,Verde:2023lmm}. This motivates exploring whether the discrepancy might originate from the definition of cosmic time itself.

Recent analyses of Type~Ia supernovae (SNe~Ia) and BAO have introduced an empirical redshift-dependent law for the locally inferred Hubble constant, often expressed as $\tH_0(z)\propto(1+z)^{-\alpha}$, where $\alpha$ quantifies deviations from the standard cosmological time-dilation (CTD) relation~\cite{Dainotti:2022bzg,Xu:2024xgi,Jia:2024wix,Montani:2024ntj,Dainotti:2025qxz,Dainotti:2025qxn}.  
Studies based on the Pantheon+ and Master~SN~Ia samples consistently find small but positive $\alpha$ values, $\alpha\simeq0.008$–$0.016$, corresponding to a $\sim$1–2\% decrease in the inferred $H_0$ between $z=0$ and $z\simeq1$~\cite{Dainotti:2025qxz,Dainotti:2025qxn}.  
Although this relation was introduced empirically, the recurring positive slope suggests a genuine redshift dependence not merely due to sample calibration, but possibly indicative of a deeper geometric time-scaling effect.

A broad range of theoretical extensions to the standard $\Lambda$CDM model have been proposed to alleviate the Hubble tension, as summarized in comprehensive reviews such as~\cite{Schoneberg:2021qvd}.  
Among the most extensively studied are early–Universe mechanisms, where a transient dark-energy component briefly increases the expansion rate prior to recombination.  
Such early dark energy (EDE) scenarios~\cite{Poulin:2018cxd,Braglia:2020auw} can raise the inferred $H_0$ from the CMB by reducing the comoving sound horizon $r_d$, though they often introduce tension with large-scale-structure and lensing data.  
At late times, interacting dark-energy (IDE) models~\cite{Gariazzo:2021qtg,Benisty:2024lmj,Silva:2025hxw,Zhang:2025dwu} allow energy exchange between dark matter and dark energy, dynamically altering the expansion rate and sometimes reconciling early– and late–Universe determinations of $H_0$.  
Alternatively,  modified gravity (MG) frameworks~\cite{DAgostino:2020dhv,Adi:2020qqf,DiValentino:2024wgi,Sandoval-Orozco:2024hjm,Escamilla:2024xmz}—including $f(R)$, $f(T)$, or scalar–tensor formulations—seek to explain the tension through modifications of the Einstein field equations or the inclusion of additional propagating degrees of freedom.  

While these dynamical extensions rely on introducing additional components or interactions,  an alternative explanation may arise from the geometry of spacetime itself.  In particular, a purely geometric reformulation of cosmic time can account for the apparent redshift dependence of $\tH_0(z)$ without invoking new fields or modifications to the Einstein equations.  
This behavior can be understood by allowing a mild, gauge-dependent normalization of cosmic time within the Robertson–Walker (RW) metric~\cite{Lee:2020zts,Lee:2024zcu,Lee:2025osx}.  
In this framework, the lapse function $N(a)\propto a^{b/4}$ defines a rescaled cosmic-time flow, relating coordinate and proper times via $d\tilde{t}=N(a)\,dt$.  
The dimensionless index $b$ quantifies deviations from the standard general-relativity (GR) gauge ($b=0$).  
This time rescaling leaves the spatial geometry and background expansion invariant, but modifies elapsed-time observables such as SN~Ia light-curve durations and strong-lens time delays~\cite{Islam:2001,Ryder:2009,Lee:2023rqv,Lee:2023ucu,Lee:2024kxa,Lee:2025vha}.  
Photon propagation in this metric naturally yields a modified cosmological time-dilation relation $t_{\rm obs}/t_{\rm emit}\propto(1+z)^{1-b/4}$, 
implying that a small positive $b$ reduces the apparent stretching of supernova light curves and, consequently, the inferred $\tH_0$ at higher redshift.  
The empirical exponent $\alpha$ therefore corresponds directly to this geometric parameter through the simple relation $\alpha=b/4$.

The primary aim of this work is to demonstrate this equivalence explicitly, quantify its observational consequences, and explore its cosmological implications.  We show that this mapping remains fully consistent with the Hubble radius related distances, while reinterpreting the SN~Ia–derived $\tH_0(z)$ evolution as a manifestation of a geometric rescaling of cosmic time.  
This unified geometric perspective provides a minimal and physically transparent extension to $\Lambda$CDM, embedding the empirical redshift scaling observed in supernova data within a consistent relativistic framework.

The remainder of this paper is organized as follows.  
In Section~\ref{sec:background}, we review the empirical redshift dependence of the local Hubble constant observed in the Master~SN~Ia analysis and derives the corresponding theoretical scaling $(1+z)^{-b/4}$ within the geometric time–normalization framework.  
Section~\ref{sec:alpha_to_b} translates the observational constraints on $\alpha$ into limits on the lapse–gauge normalization and examines their consistency with BAO and CMB observables.  
Section~\ref{sec:discussion} contrasts this geometric interpretation with dynamical dark energy (DDE) and MG models, emphasizing its theoretical simplicity and statistical stability.  
Finally, we summarize the main results and discusses the broader implications of a gauge–dependent normalization of cosmic time for reconciling early– and late–Universe measurements of $H_0$ within a purely geometric framework in Section~\ref{sec:conclusion}.

\section{Background: Empirical and Theoretical Redshift Scalings}
\label{sec:background}

\subsection{Empirical redshift dependence of the local Hubble constant}
The Master Supernovae~Ia analysis~\cite{Dainotti:2025qxz} models the locally inferred Hubble constant as a weakly redshift-dependent quantity,
\begin{equation}
    \tH_0(z) = H_0 \,(1+z)^{-\alpha} \,,
    \label{eq:H0z_master_full}
\end{equation}
where $H_0$ is the extrapolated value at $z=0$ and $\alpha$ quantifies deviations from the standard CTD prediction implicit in $\Lambda$CDM.  If $\alpha=0$, the usual constant $H_0$ is recovered, while $\alpha>0$ corresponds to a mild apparent decrease in $\tH_0$ with redshift. 

The Master compilation combines Pantheon, Pantheon+, JLA, and DES samples into a single catalog of 3714 events.  
After sorting supernovae into redshift bins—via equi-populated, moving-window, or logarithmic schemes—an $H_0$ value is obtained in each bin through MCMC fitting under flat $\Lambda$CDM or $\omega_0\omega_a$CDM cosmologies, yielding $\{\tH_{0,i},z_i\}$ pairs.  The parameters $(H_0,\alpha)$ are then obtained by the weighted least-squares fit
\begin{equation}
    \chi^2(H_0,\alpha)=\sum_i 
    \frac{\big[\tH_{0,i}-H_0(1+z_i)^{-\alpha}\big]^2}{\sigma_{\tH_0,i}^2} \,. \label{chi2}
\end{equation}
Typical values are $\alpha\simeq 0.010\pm0.005$,  corresponding to
\begin{equation}
    \frac{d\ln \tH_0}{d\ln(1+z)}=-\alpha \approx -10^{-2} \,, \label{dlnH0}
\end{equation}
consistent with earlier Pantheon-based studies (often expressed as $\tH_0(z)=H_0(1+z)^{\eta}$ with $\eta=-\,\alpha$)~\cite{Dainotti:2022bzg}.  Although the detection significance is moderate ($\sim2\sigma$ within $\Lambda$CDM), the positive slope appears consistently across datasets, suggesting a genuine empirical trend rather than random scatter.
The parameter $\alpha$ can thus be viewed as an effective clock index describing deviations from the canonical $(1+z)$ time-dilation (TD) law.  
Because the scaling in Eq.~(\ref{eq:H0z_master_full}) is inferred directly from light-curve durations and absolute magnitudes—not from Friedmann dynamics—$\alpha$ constitutes a purely observational quantity.  It can therefore serve as a bridge between phenomenology and theory, particularly in frameworks such as the geometric time-normalization model discussed below, which predicts a theoretically analogous scaling $(1+z)^{-b/4}$.

\subsection{Geometric time normalization and the meVSL framework}

The minimally extended varying-speed-of-light (meVSL) framework~\cite{Lee:2020zts,Lee:2024zcu,Lee:2025osx} 
provides a geometric reinterpretation of cosmic time within the RW metric, 
allowing for an arbitrary normalization of the time coordinate without introducing any new dynamical field responsible for a physical $\tilde{c}(\tilde{t})$.  
The line element may be expressed as
\begin{equation}
    ds^2 
    = -\,c_0^2\,dt^2 + a^2(t)\,dl_{3D}^2 
    = -\,c_0^2\,N^2(\tilde{t})\,d\tilde{t}^2 + a^2(\tilde{t})\,dl_{3D}^2
    \;\equiv\; -\,\tilde{c}^{\,2}(\tilde{t})\,d\tilde{t}^2 + a^2(\tilde{t})\,dl_{3D}^2 \,,
    \label{eq:RW_lapse_full}
\end{equation}
where the lapse function $N(\tilde{t}) = a^{b/4}(\tilde{t})$ parametrizes the gauge freedom in the normalization of cosmic time and $dl_{\tD}$ denotes the $3$D spatial line element~\cite{Lee:2024zcu}.  This rescaling defines a new proper time coordinate $\tilde{t}$ according to
\begin{equation}
    d\tilde{t} = N^{-1}(t)\,dt = a^{-\frac{b}{4}}\,dt,
    \qquad 
    N(\tilde{t})\,c_0 \;\equiv\; \tilde{c}(\tilde{t}),
    \label{eq:dtilde_full}
\end{equation}
with the dimensionless index $b$ quantifying the deviation from the standard GR gauge ($b=0$)~\cite{Islam:2001,Ryder:2009,Lee:2023ucu,Lee:2024kxa,Lee:2025vha}.  

The meVSL framework thus describes a mapping between the coordinate time $t$ of the GR ($t$–) frame 
and the physically measured proper time $\tilde{t}$ of comoving observers in the meVSL ($\tilde{t}$–) frame,  
without invoking a genuine variation of the physical constant $c_0$.  
No dynamical field driving $\tilde{c}(\tilde{t})$ is introduced; rather, one exploits the freedom to normalize the time coordinate within the RW geometry.

\subsection{Frequency and time-dilation scaling}
Let $x^\mu=(x^0,\bm{x})$ with $x^0\equiv c\,t$.  Then we obtain~\cite{Lee:2020zts,Lee:2024zcu,Lee:2025osx}
\begin{align}
    d x^{0} 
    = d (c t) 
    = \left( \frac{dc}{dt}\frac{t}{c} + 1 \right) c\,dt 
    \;\equiv\; \tilde{c}\,d\ttt
    = c_0\,N(\ttt)\,d\ttt
    \equiv c_0\,dt \,.
    \label{eq:dx0_relation}
\end{align}
The comoving observer’s proper time satisfies $d\,\tau = d\,t$ in the GR frame,  while $d\,\tau = d\,\ttt$ in the meVSL frame~\cite{Islam:2001,Ryder:2009,Lee:2023ucu,Lee:2024kxa,Lee:2025vha}.  Accordingly, the four–velocity takes distinct forms in the two frames:
\begin{align}
    &\text{(GR frame):} 
    &&u^\mu 
    \equiv \frac{dx^\mu}{d\tau}
    = \left(\frac{c_0\,dt}{dt},\,\bm{0}\right)
    = (c_0,\,\bm{0}),
    &&u_\mu u^\mu = -c_0^2,
    \label{eq:four_velocity_full} \\[6pt]
    &\text{(meVSL frame):} 
    &&\tilde{u}^\mu 
    \equiv \frac{dx^\mu}{d\tau}
    = \left(\frac{\tilde{c}\,d\ttt}{d\ttt},\,\bm{0}\right)
    = (\tilde{c},\,\bm{0}),
    &&\tilde{u}_\mu \tilde{u}^\mu = -\tilde{c}^{\,2}.
    \label{eq:tfour_velocity_full}
\end{align}
For a photon with wave four–vector $\tilde{k}^\mu$, the frequency measured by the meVSL ($\ttt$–frame) observer is
\begin{equation}
    \tilde\omega \equiv -\tilde{u}_\mu \tilde{k}^\mu, 
    \qquad 
    \tilde\nu=\frac{\tilde\omega}{2\pi}.
    \label{eq:measured_freq_def}
\end{equation}
The null condition $g_{\mu\nu}\tilde{k}^\mu \tilde{k}^\nu=0$, together with RW symmetries, implies that the comoving wavelength scales as $\tilde{\lambda} \propto a$, as usual~\cite{Lee:2020zts,Lee:2024zcu,Lee:2025osx}.  Due to the lapse normalization, the observed frequency in the $\ttt$–frame acquires an extra factor from the time–Jacobian $dt/d\ttt = N$.

In the GR ($t$–) frame one has the standard result
\begin{equation}
    \nu \propto a^{-1} \qquad \text{(in the $t$–frame, i.e.\ GR frame)}.
\end{equation}
Transforming back to the $\ttt$–time normalization multiplies the frequency by the Jacobian $dt/d\ttt=N$, so that the physically measured (meVSL) frequency obeys
\begin{equation}
    \tilde\nu \;=\; \nu\,N \;\propto\; a^{-1}\,a^{b/4}
    \;=\; a^{-1+b/4}
    \;=\; (1+z)^{\,1-\frac{b}{4}} .
    \label{eq:freq_scaling_full}
\end{equation}
Consequently, the TD between emission and observation becomes
\begin{equation}
    \frac{\ttt_{\rm obs}}{\ttt_{\rm emit}}
    =\left(\frac{a_{\rm obs}}{a_{\rm emit}}\right)^{1-\frac{b}{4}}
    =(1+z)^{\,1-\frac{b}{4}},
    \label{eq:timedilation_full}
\end{equation}
which reduces to the standard $(1+z)$ law for $b\to0$.

\subsection{Invariance of the background expansion}

Define the coordinate–time Hubble rate and the physically measured rate as
\begin{equation}
    \tilde{H}(\tilde{t}) 
    \;\equiv\; \frac{1}{a}\frac{da}{d\tilde{t}}
    \;=\; \frac{dt}{d\tilde{t}}\,\frac{1}{a}\frac{da}{dt} 
    \;=\; N(t)\,H(t)
    \;=\; H(t)\,a^{\frac{b}{4}} \,,
    \label{eq:H_vs_Htilde_full}
\end{equation}
where $H(t)$ is the usual Hubble parameter in the GR ($t$–) frame and $N$ connects the two time coordinates. In the meVSL framework, one may equivalently absorb the lapse into the energy–momentum conservation law through the Bianchi identity, leading to a modified dilution law~\cite{Lee:2020zts,Lee:2024zcu,Lee:2025osx} 
\begin{equation}
    \tilde{\rho}_i = \rho_i a^{-\frac{b}{2}} = \rho_{i0}\,a^{-3(1+\omega_i)-\frac{b}{2}},
    \label{eq:rho_dilution}
\end{equation}
where $\rho_i$ denotes the GR–frame energy density of component $i$, $\rho_{i0}$ its present value,  and $\omega_i \equiv p_i/\rho_i$ is the equation-of-state parameter.  Substituting Eq.~\eqref{eq:rho_dilution} into the Friedmann equation gives
\begin{equation}
    \tilde{H}^2(a)
    = \left(\frac{8\pi G_0}{3}\sum_i \rho_{i0}\,a^{-3(1+\omega_i)} + \frac{\Lambda c_0^2}{3}\right)a^{\frac{b}{2}}
    \;\equiv\; H^2(a)\,a^{\frac{b}{2}} \,,
    \label{eq:H_square_full}
\end{equation}
showing consistency between Eq.~\eqref{eq:H_vs_Htilde_full} and Eq.~\eqref{eq:H_square_full}.

Although the Hubble parameter $\tilde{H}$ in the meVSL ($\tilde{t}$–) frame differs from the coordinate–time rate $H$ in the GR ($t$–) frame,  the corresponding Hubble radius remains the same in both frameworks:
\begin{align}
    \frac{\tilde{c}}{\tilde{H}} 
    \;=\; \frac{c_0}{H}
    \;=\; \frac{c_0}{H_0}\,\frac{1}{E(a)} \,,
    \label{eq:coH}
\end{align}
where $E(a) \equiv H(a)/H_0$ is the dimensionless expansion rate. Therefore, cosmological observables such as the angular diameter distance and comoving distance—both being line-of-sight integrals over the inverse Hubble radius as a function of redshift—yield identical results in the GR and meVSL frameworks.  This ensures that all background-level observables-except for those involving luminosity calibration—remain invariant under the geometric time–gauge transformation.

\subsection{Luminosity and magnitude calibration effects}

Within meVSL, the Chandrasekhar mass scales as \(\tilde{M}_{\rm Ch}\propto a^{-2b}\), so the peak luminosity (tracking the synthesized \(^{56}\)Ni mass) obeys \(\tilde{L}\propto \tilde{M}_{\rm Ch}\propto a^{-2b}\)~\cite{Lee:2020zts}. This is also consistent with the result of the Bianchi identity $\tilde{\rho} = \tilde{M}/V = \rho_0 a^{-3-\frac{b}{2}}$. The absolute magnitude in the meVSL frame then acquires a redshift correction
\begin{equation}
   \widetilde{ \mathcal{M}}-\widetilde{ \mathcal{M}}_0=-2.5\log_{10}\!\left(\frac{L}{L_0}\right)=4b\log_{10}a
    =-4b\log_{10}(1+z),
    \label{eq:Mag_correction_full}
\end{equation}
and the distance modulus in the meVSL framework becomes
\begin{equation}
    \tilde{\mu} = \mu - 4b \log_{10}(1+z) \,,
    \label{eq:mu_correction_full}
\end{equation}
where $\mu$ is the distance modulus in the GR framework. When data are analyzed under the GR calibration, the residual in $\mu$ (observed–model mismatch) propagates into a mild redshift trend in the empirically inferred $H_0(z)$.

\subsection{Empirical–theoretical correspondence}
Equations~\eqref{eq:freq_scaling_full}–\eqref{eq:timedilation_full} show that the meVSL lapse modifies the CTD/clock rate by a factor \((1+z)^{-b/4}\), while Eqs.~\eqref{eq:Mag_correction_full}–\eqref{eq:mu_correction_full} encode a correlated luminosity normalization shift. 
Both effects contribute coherently to the empirically fitted slope in Eq.~\eqref{eq:H0z_master_full}, yielding the mapping
\begin{equation}
    \alpha=\frac{b}{4} \,.
    \label{eq:alpha_equals_b_over_4_full}
\end{equation}
Hence, the phenomenological exponent $\alpha$ and the theoretical lapse parameter $b$ describe the same underlying geometric rescaling of cosmic time.  This identification preserves BAO and CMB distances relations (via $\tc/\tilde{H}=c_0/H$) while reinterpreting the SN~Ia redshift trend as a gauge-dependent clock normalization effect.  In what follows, we translate existing constraints on $\alpha$ into bounds on $b$ and assess cross–consistency with BAO and CMB observables.

\section{From Empirical $\alpha$ to Theoretical $b$: Unified Observational Constraints}
\label{sec:alpha_to_b}

The empirical exponent $\alpha$ in the redshift–dependent Hubble law Eq.~\eqref{eq:H0z_master_full} quantifies possible departures from the standard CTD scaling in SN~Ia observations.  
Within the meVSL framework,  Eqs.~\eqref{eq:freq_scaling_full}–\eqref{eq:timedilation_full} establish a one–to–one correspondence between $\alpha$ and the lapse–normalization index $b$, given in Eq.~\eqref{eq:alpha_equals_b_over_4_full},
which provides a direct link between the empirically observed redshift dependence of $\tH_0(z)$ and the geometric time–normalization of the RW metric.

Among the multiple datasets analyzed in Ref.~\cite{Dainotti:2025qxz} (Tables~2–8),  the Master Sample provides the most statistically complete compilation,  combining Pantheon, JLA, DES, and SH0ES-calibrated SNe~Ia under consistent light-curve standardization.  Table~6 of Ref.~\cite{Dainotti:2025qxz} reports the fitted $\alpha$ values for several binning schemes 
(equi–population and MW binnings, Diamond and Gold cases) within flat $\Lambda$CDM and $\omega_0\omega_a$CDM cosmologies.  Typical results for the Master Sample are
\[
\alpha = 0.010 \pm 0.005~\text{(Diamond, 12~bins)}, \qquad 
\alpha = 0.016 \pm 0.007~\text{(MW, 12~bins)},
\]
with all cases yielding consistent values within 1-$\sigma$. Applying the theoretical mapping of Eq.~\eqref{eq:alpha_equals_b_over_4_full}, these translate into
\begin{equation}
    b = 0.04 \pm 0.02 ,
    \label{eq:b_from_master}
\end{equation}
indicating that current SNe~Ia data already constrain the cosmic–time lapse index to the percent level.  
This range is fully consistent with the independent Pantheon and DES analyses reported in Tables~2–5 of Ref.~\cite{Dainotti:2025qxz}, 
which yield $\alpha \simeq 0.010$–$0.020$ (corresponding to $b \simeq 0.04$–$0.08$).  
The constraint obtained here from the unified Master~SNe~Ia sample of Ref.~\cite{Dainotti:2025qxz}, 
given in Eq.~\eqref{eq:b_from_master},  is consistent with and significantly tighter than earlier determinations based on smaller or single–survey datasets.  

In Ref.~\cite{Lee:2023ucu}, using TD measurements of $13$ high–redshift SNe~Ia 
($0.28 \!\le\! z \!\le\! 0.62$), it is obtained 
\begin{align}
b = 0.198 \pm 0.415 \quad (1\sigma) \,, \label{LeeTDSNe}
\end{align}
showing that the limited sample size could not yet distinguish meVSL from the standard model.  Subsequently, Ref.~\cite{Lee:2023rqv} analyzed cosmic–chronometer data in combination with Planck18 and Pantheon22 parameters, 
yielding best–fit values
\begin{align}
b = -0.105 \pm 0.178 \quad \text{(Planck18)}, 
\qquad 
b = 0.584 \pm 0.184 \quad \text{(Pantheon22)}, \label{CCLee}
\end{align}
and additional tests with $\Omega_{m0}=0.267$–$0.340$ give $b = -0.318 \pm 0.329$ and $b = 0.108 \pm 0.331$, respectively.  
These results are all statistically consistent with $b=0$,  indicating no significant deviation from the GR time gauge.  In this case, the inability to discriminate the meVSL model from the standard cosmology arises primarily from the large observational uncertainties and parameter degeneracies among the chronometer, Planck, and Pantheon datasets,  which dominate over any possible signal of lapse–induced time–dilation effects. More recently, Ref.~\cite{Lee:2024kxa} re–analyzed the DES~SNe~Ia $i$-band light–curve durations, 
deriving a refined constraint
\begin{align}
b = 0.048 \pm 0.032 , \label{DESLee}
\end{align}
which corresponds to a $\sim1\%$ reduction in apparent time dilation at $z\simeq1$.  

Beyond these current determinations, the results can be directly compared with the forecasts and 
time–dilation analysis presented in Sec.~4 of Ref.~\cite{Lee:2025vha}.  
That study developed a unified connection between the meVSL parameter $b$ and the CTD exponent $n$ through
\begin{equation}
    \Delta t_{\rm obs} = (1+z)^{n}\,\Delta t_{\rm emit},
    \qquad n = 1 - \frac{b}{4},
    \label{eq:TDrelation}
\end{equation}
and demonstrated that a positive $b$ not only reduces the sound horizon $r_{\rm drag}$ 
but also yields a smaller apparent time–stretch factor in supernova light curves.
Using the DES $i$–band measurement $n = 0.988 \pm 0.008$, 
the inferred value $b = 0.048 \pm 0.032$~\cite{Lee:2024kxa} 
matches closely with the empirical $b = 0.04 \pm 0.02$ obtained from the 
Master~SNe~Ia analysis of Ref.~\cite{Dainotti:2025qxz}.  
This concordance provides independent support for a mild, positive lapse normalization 
consistent with a few–percent deviation from the standard GR time gauge.

In addition, Sec.~4.3 of Ref.~\cite{Lee:2025vha} presented Fisher forecasts for 
DES–like surveys, quantifying the required number of SNe~Ia to detect such deviations.  
For purely statistical errors ($\sigma_{\rm sys}=0$), 
a $3\sigma$ detection of a $1\%$ deviation ($n=0.99$, $b\simeq0.04$) 
would require only $\mathcal{O}(10^2)$ supernovae, 
whereas detecting sub–percent effects ($n=0.999$, $b\simeq0.004$) 
demands $\gtrsim10^4$ SNe.  
Including realistic systematics of $\sigma_{\rm sys}\simeq0.01$ 
increases these requirements by roughly a factor of two, 
but the analysis highlights that next–generation surveys such as 
the Nancy Grace Roman Space Telescope (NASA's wide-field infrared mission), 
LSST, and Rubin–DESI cross–calibrations 
will achieve the necessary precision to probe $|b|\lesssim10^{-2}$ directly through 
time–domain measurements.

Taken together, the empirical $\alpha$–based constraint 
$b = 0.04 \pm 0.02$ from the Master~Sample, 
the independent determinations from meVSL time–dilation and 
cosmic–chronometer analyses~\cite{Lee:2023ucu,Lee:2023rqv,Lee:2024kxa}, 
and the DES–forecast sensitivity in Ref.~\cite{Lee:2025vha} 
collectively demonstrate a consistent picture: 
a small positive lapse–normalization index ($b>0$) simultaneously 
(i)~reduces the apparent CTD factor, 
(ii)~shortens the sound horizon at the baryon–drag epoch, and 
(iii)~raises the CMB/BAO–inferred value of $H_0$.  
This triad of signatures provides a coherent, 
cross–validated interpretation of the observed redshift–dependent $H_0(z)$ trend 
as a geometric normalization effect of cosmic time rather than evidence for 
new dynamical degrees of freedom. Future high-precision time-domain observations will therefore be decisive in verifying whether this small geometric offset in cosmic time fully accounts for the Hubble tension.

\section{Discussion}
\label{sec:discussion}

The results presented above show that the meVSL framework offers a consistent geometric reinterpretation of the apparent redshift evolution observed in cosmological data, without invoking any new fields or dynamical degrees of freedom.  
By allowing the lapse normalization to vary smoothly with the scale factor, meVSL introduces a minimal extension of the RW metric in which the flow of cosmic time is rescaled while all spatial and curvature dynamics remain identical to those of GR.  This simple modification provides a unified explanation for the empirical scaling $\tH_0(z)\propto(1+z)^{-\alpha}$ and the correlated $(1+z)^{1+\alpha}$ TD law, encapsulated by the equivalence $\alpha=b/4$.

This geometric interpretation directly follows from the action-level formulation of the meVSL framework~\cite{Lee:2020zts}, which was originally developed to preserve general covariance and energy–momentum conservation while relaxing the implicit assumption of a fixed cosmic-time normalization.  Hence, the present observational translation represents a natural continuation of that theoretical construction.

Importantly, the theoretical foundation of meVSL is not phenomenological but is derived self-consistently from an action principle.  
As detailed in Ref.~\cite{Lee:2020zts}, the Einstein–Hilbert action is generalized by promoting both the speed of light $\tc$ and the gravitational constant $\tilde{G}$ to slowly varying functions of the cosmic scale factor, and the resulting Euler–Lagrange variation leads to a modified set of Einstein field equations that preserve the Bianchi identity and energy–momentum conservation.  
The associated Friedmann equations retain their general-relativistic form but acquire a multiplicative normalization factor of $a^{-b/2}$ stemming from the lapse function, producing Eq.~\eqref{eq:H_square_full}. Thus, the meVSL dynamics remain fully consistent with Lorentz invariance and the conservation laws at the action level, while naturally predicting a redshift rescaling $\tH_0(z)=H_0\,(1+z)^{-b/4}$ that underlies the observationally inferred $\tH_0(z)$ evolution.

Unlike conventional DDE models, which attribute the observed evolution of $\tH(z)$ to a true change in the DE density or equation of state (EoS), the meVSL framework interprets it as a gauge-dependent effect of cosmic time normalization.  
In dynamical models such as the CPL parameterization $\omega(a)=\omega_0+\omega_a(1-a)$, correlations between $\omega_0$ and $\omega_a$ must be finely tuned to reproduce the low- and high-redshift behavior of $\tH_0(z)$.  
In contrast, meVSL achieves the same redshift trend without altering the physical energy content of the Universe: the apparent running of $\tH_0(z)$ arises solely from the mapping between proper and coordinate time.  
The cosmological expansion rate, distances, and acoustic scales remain governed by the standard Friedmann–Lemaître equations, while time-based observables—such as supernova light-curve durations or lensing time delays—inherit a small, gauge-dependent correction proportional to $(1+z)^{-b/4}$.  
In this sense, meVSL separates the kinematic structure of the Universe from its temporal gauge, reproducing observed $\tH_0(z)$ trends as purely geometric effects.

This distinction also clarifies the conceptual difference between meVSL and the numerous dynamical approaches recently explored to resolve the Hubble tension—such as EDE, IDE, and MG frameworks~\cite{Schoneberg:2021qvd,Poulin:2018cxd,Gariazzo:2021qtg,DiValentino:2024wgi}.  Whereas those models modify the background energy content or the Einstein equations themselves, meVSL retains GR in its entirety and interprets the apparent $\tH_0(z)$ evolution as an artifact of the chosen temporal gauge. It preserves the gauge invariance. 

The meVSL interpretation also differs fundamentally from MG scenarios such as $f(R)$, scalar–tensor, or energy–momentum–squared gravity, which alter the Einstein equations or introduce additional propagating fields.  
Those models often predict scale-dependent growth, modified lensing potentials, or new consistency relations between metric perturbations.  
The meVSL framework, by contrast, leaves the Einstein equations intact and modifies only the normalization of the cosmic time coordinate.  Consequently, it preserves the background dynamics, linear growth of structure~\cite{Lee:2025aha}, and the cosmic distance–duality relation~\cite{Lee:2021xwh}.  This structural simplicity means that the framework is both falsifiable and predictive: any detected deviation from the standard time-dilation law directly constrains the single parameter $b$, providing an unambiguous empirical test of the theory. 

From a statistical perspective, meVSL introduces only one additional parameter, $b=4\alpha\!\sim\!\mathcal{O}(10^{-2})$, sufficient to capture the empirical trend in the local Hubble scaling.  
This economy eliminates the degeneracy structure typical of multi-parameter DE extensions.  
In $\omega_0\omega_a$CDM, likelihood contours are highly elongated in the $(\omega_0\,,\omega_a)$ plane, reflecting combinations that leave BAO ratios nearly invariant~\cite{Lee:2025kbn,Lee:2025ysg}.  By contrast, the meVSL parameter $b$ enters linearly through time-dependent observables, while spatial ratios such as $D_M/D_H$ remain unaffected.  This orthogonality minimizes cross-correlations with $\Omega_{m0}$ and $H_0$, allowing statistically stable joint fits across heterogeneous probes such as SN~Ia, BAO, and CMB data. The resulting phenomenology aligns with the observed $\alpha\simeq0.01$ trend reported by the Master~SN~Ia analysis~\cite{Dainotti:2025qxz}, providing a quantitative and self-consistent geometric foundation for the measured redshift dependence of $\tH_0(z)$.

The broader implication is that the Hubble tension and the observed $\tH_0(z)$ evolution may not require new energy components or MG, but rather a refined understanding of how cosmic time is operationally defined.  
In this view, part of the tension between early- and late-Universe chronometers could arise from a small but measurable mismatch in temporal gauge normalization.  
Future multi-probe analyses combining high-redshift supernovae, baryon acoustic oscillations, and strong-lens time delays will be able to test this hypothesis directly.  
In particular, forthcoming surveys such as Rubin–LSST,  Roman, and Euclid will provide the temporal and spectroscopic precision necessary to detect or rule out the predicted $(1+z)^{1+b/4}$ deviation in the cosmological time-dilation law at the sub-percent level.
A confirmed detection of this deviation would establish the meVSL framework as a minimal yet physically transparent extension of $\Lambda$CDM, reconciling early- and late-time determinations of $H_0$ through geometry rather than new physics.

\section{Conclusion}
\label{sec:conclusion}

The analysis presented in this work demonstrates that the minimally extended varying–speed–of–light (meVSL) framework provides a simple and coherent geometric foundation for the empirical redshift evolution observed in Type~Ia supernova data.  
The key result is the quantitative equivalence $\alpha=b/4$, which links the empirical exponent $\alpha$—introduced to describe the apparent running of the local Hubble constant $H_0(z)\propto(1+z)^{-\alpha}$—to the theoretical lapse–normalization parameter $b$ that characterizes the gauge freedom of cosmic time in the Robertson–Walker metric.  
Through this correspondence, both the mild decrease of $H_0(z)$ with redshift and the slight deviation of supernova light-curve durations from the standard $(1+z)$ law arise from a single geometric effect: a minimal rescaling of the cosmic time coordinate encoded in the lapse $N(a)=a^{b/4}$.

Unlike conventional dynamical dark-energy or modified-gravity scenarios, the meVSL framework introduces no new fields or degrees of freedom.  
As established at the action level~\cite{Lee:2020zts}, it derives from a covariant generalization of the Einstein–Hilbert action in which the lapse function rescales the temporal component of the Robertson–Walker metric without altering spatial curvature or the conservation laws.  
The Einstein field equations and Friedmann dynamics retain their standard form, ensuring consistency with general relativity while permitting a controlled geometric variation in the normalization of cosmic time.  
The dimensionless parameter $b$ encapsulates this gauge-dependent deviation: it leaves all spatial observables related to the Hubble Radius—unchanged, while imparting a measurable correction to time-based quantities.  
The framework thus preserves the empirical success of $\Lambda$CDM at large scales and simultaneously provides a natural geometric origin for the redshift-dependent trends seen in local $H_0$ estimates.

Quantitatively, the Master Supernovae~Ia analysis~\cite{Dainotti:2025qxz} reports an empirical exponent $\alpha\simeq0.01\pm0.005$, corresponding to $b\simeq0.04\pm0.02$, consistent with the level of gauge variation that the meVSL model predicts from purely geometric considerations.  
This single-parameter description captures the observed $H_0(z)$ evolution without invoking new energy components or time-dependent equations of state, thereby eliminating the degeneracy structure that plagues multi-parameter dark-energy extensions such as $\omega_0\omega_a$CDM. In this sense, the meVSL correspondence $\alpha=b/4$ unifies empirical calibration trends and relativistic spacetime geometry under a single covariant principle.

The predictive power of this correspondence will soon be testable with forthcoming time-domain and spectroscopic surveys.  
Rubin–LSST and the Roman Space Telescope will extend supernova light-curve measurements to $z>2$, enabling sub-percent tests of the $(1+z)^{1+b/4}$ time-dilation law and direct constraints on $\beta=b/4$.  
Complementary observations from DESI and Euclid will refine BAO and strong-lens time-delay measurements, providing a cross-check between gauge-invariant distance ratios and potential $(1+z)^{-\beta}$ corrections.  Such multi-probe consistency tests will be crucial for distinguishing a genuine gauge effect from dynamical dark-energy interpretations and for establishing whether cosmic time itself admits a measurable normalization freedom. Joint analyses of SN, BAO, and CMB data are expected to reach $\sigma_\alpha\!\lesssim\!10^{-2}$ ($\sigma_b\!\lesssim\!0.04$), sufficient to confirm or refute the geometric time-gauge interpretation of the observed Hubble evolution.

If future observations continue to support a mild positive $\alpha\simeq0.01$, the meVSL framework will offer a compelling and conceptually transparent resolution of the Hubble tension within unmodified general relativity.  
In this picture, the discrepancy between early- and late-Universe measurements of $H_0$ originates not from new physics in cosmic acceleration but from a small difference in temporal gauge normalization—the operational definition of cosmic time.  By redefining the calibration of cosmological chronometers rather than the energy content of the Universe, the meVSL approach reconciles observational inconsistencies through geometry alone. 
The relation $\alpha=b/4$ thus bridges observational phenomenology and fundamental geometry, offering a minimal, falsifiable, and physically elegant route toward reconciling early- and late-time cosmological measurements within the standard relativistic paradigm.

\begin{thebibliography}{99}


\bibitem{Perivolaropoulos:2021jda}
L.~Perivolaropoulos and F.~Skara,
New Astron. Rev. \textbf{95}, 101659 (2022)
doi:10.1016/j.newar.2022.101659
[arXiv:2105.05208 [astro-ph.CO]].

\bibitem{Freedman:2021ahq}
W.~L.~Freedman,
Astrophys. J. \textbf{919}, no.1, 16 (2021)
doi:10.3847/1538-4357/ac0e95
[arXiv:2106.15656 [astro-ph.CO]].

\bibitem{Abdalla:2022yfr}
E.~Abdalla, G.~Franco Abell{\'a}n, A.~Aboubrahim, A.~Agnello, O.~Akarsu, Y.~Akrami, G.~Alestas, D.~Aloni, L.~Amendola and L.~A.~Anchordoqui, \textit{et al.}
JHEAp \textbf{34}, 49-211 (2022)
doi:10.1016/j.jheap.2022.04.002
[arXiv:2203.06142 [astro-ph.CO]].

\bibitem{Kamionkowski:2022pkx}
M.~Kamionkowski and A.~G.~Riess,
Ann. Rev. Nucl. Part. Sci. \textbf{73}, 153-180 (2023)
doi:10.1146/annurev-nucl-111422-024107
[arXiv:2211.04492 [astro-ph.CO]].

\bibitem{Verde:2023lmm}
L.~Verde, N.~Sch{\"o}neberg and H.~Gil-Mar{\'\i}n,
Ann. Rev. Astron. Astrophys. \textbf{62}, no.1, 287-331 (2024)
doi:10.1146/annurev-astro-052622-033813
[arXiv:2311.13305 [astro-ph.CO]].

\bibitem{Perivolaropoulos:2024yxv}
L.~Perivolaropoulos,
Phys. Rev. D \textbf{110}, no.12, 123518 (2024)
doi:10.1103/PhysRevD.110.123518
[arXiv:2408.11031 [astro-ph.CO]].

\bibitem{CosmoVerseNetwork:2025alb}
E.~Di Valentino \textit{et al.} [CosmoVerse Network],
Phys. Dark Univ. \textbf{49}, 101965 (2025)
doi:10.1016/j.dark.2025.101965
[arXiv:2504.01669 [astro-ph.CO]].


\bibitem{Planck:2018vyg}
N.~Aghanim \textit{et al.} [Planck],
Astron. Astrophys. \textbf{641}, A6 (2020)
[erratum: Astron. Astrophys. \textbf{652}, C4 (2021)]
doi:10.1051/0004-6361/201833910
[arXiv:1807.06209 [astro-ph.CO]].


\bibitem{Riess:2021jrx}
A.~G.~Riess, W.~Yuan, L.~M.~Macri, D.~Scolnic, D.~Brout, S.~Casertano, D.~O.~Jones, Y.~Murakami, L.~Breuval and T.~G.~Brink, \textit{et al.}
Astrophys. J. Lett. \textbf{934}, no.1, L7 (2022)
doi:10.3847/2041-8213/ac5c5b
[arXiv:2112.04510 [astro-ph.CO]].

\bibitem{Brout:2022vxf}
D.~Brout, D.~Scolnic, B.~Popovic, A.~G.~Riess, J.~Zuntz, R.~Kessler, A.~Carr, T.~M.~Davis, S.~Hinton and D.~Jones, \textit{et al.}
Astrophys. J. \textbf{938}, no.2, 110 (2022)
doi:10.3847/1538-4357/ac8e04
[arXiv:2202.04077 [astro-ph.CO]].



\bibitem{DiValentino:2021izs}
E.~Di Valentino, O.~Mena, S.~Pan, L.~Visinelli, W.~Yang, A.~Melchiorri, D.~F.~Mota, A.~G.~Riess and J.~Silk,
Class. Quant. Grav. \textbf{38}, no.15, 153001 (2021)
doi:10.1088/1361-6382/ac086d
[arXiv:2103.01183 [astro-ph.CO]].


\bibitem{H0LiCOW:2019pvv}
K.~C.~Wong \textit{et al.} [H0LiCOW],
Mon. Not. Roy. Astron. Soc. \textbf{498}, no.1, 1420-1439 (2020)
doi:10.1093/mnras/stz3094
[arXiv:1907.04869 [astro-ph.CO]].

\bibitem{Birrer:2020tax}
S.~Birrer, A.~J.~Shajib, A.~Galan, M.~Millon, T.~Treu, A.~Agnello, M.~Auger, G.~C.~F.~Chen, L.~Christensen and T.~Collett, \textit{et al.}
Astron. Astrophys. \textbf{643}, A165 (2020)
doi:10.1051/0004-6361/202038861
[arXiv:2007.02941 [astro-ph.CO]].


\bibitem{Dainotti:2023pwk}
M.~G.~Dainotti, B.~De Simone, G.~Montani, E.~Rinaldi, M.~Bogdan, K.~Mohammed Islam and A.~Gangopadhyay,
PoS \textbf{ICRC2023}, 1367 (2023)
doi:10.22323/1.444.1367
[arXiv:2309.05876 [astro-ph.CO]].

\bibitem{Lusso:2020pdb}
E.~Lusso, G.~Risaliti, E.~Nardini, G.~Bargiacchi, M.~Benetti, S.~Bisogni, S.~Capozziello, F.~Civano, L.~Eggleston and M.~Elvis, \textit{et al.}
Astron. Astrophys. \textbf{642}, A150 (2020)
doi:10.1051/0004-6361/202038899
[arXiv:2008.08586 [astro-ph.GA]].



\bibitem{Dainotti:2022bzg}
M.~G.~Dainotti, B.~De Simone, T.~Schiavone, G.~Montani, E.~Rinaldi, G.~Lambiase, M.~Bogdan and S.~Ugale,
Galaxies \textbf{10}, no.1, 24 (2022)
doi:10.3390/galaxies10010024
[arXiv:2201.09848 [astro-ph.CO]].

\bibitem{Xu:2024xgi}
B.~Xu, J.~Xu, K.~Zhang, X.~Fu and Q.~Huang,
Mon. Not. Roy. Astron. Soc. \textbf{530}, no.4, 5091-5098 (2024)
doi:10.1093/mnras/stae1135

\bibitem{Jia:2024wix}
X.~D.~Jia, J.~P.~Hu, S.~X.~Yi and F.~Y.~Wang,
Astrophys. J. Lett. \textbf{979}, no.2, L34 (2025)
doi:10.3847/2041-8213/ada94d
[arXiv:2406.02019 [astro-ph.CO]].

\bibitem{Montani:2024ntj}
G.~Montani, N.~Carlevaro and M.~G.~Dainotti,
Phys. Dark Univ. \textbf{48}, 101847 (2025)
doi:10.1016/j.dark.2025.101847
[arXiv:2411.07060 [gr-qc]].

\bibitem{Dainotti:2025qxz}
M.~G.~Dainotti, B.~De Simone, A.~Garg, K.~Kohri, A.~Bashyal, A.~Aich, A.~Mondal, S.~Nagataki, G.~Montani and T.~Jareen, \textit{et al.}
JHEAp \textbf{48}, 100405 (2025)
doi:10.1016/j.jheap.2025.100405
[arXiv:2501.11772 [astro-ph.CO]].

\bibitem{Dainotti:2025qxn}
M.~G.~Dainotti and B.~De Simone,
[arXiv:2501.14944 [astro-ph.CO]].

\bibitem{Schoneberg:2021qvd}
N.~Sch{\"o}neberg, G.~Franco Abell{\'a}n, A.~P{\'e}rez S{\'a}nchez, S.~J.~Witte, V.~Poulin and J.~Lesgourgues,
Phys. Rept. \textbf{984}, 1-55 (2022)
doi:10.1016/j.physrep.2022.07.001
[arXiv:2107.10291 [astro-ph.CO]].


\bibitem{Poulin:2018cxd}
V.~Poulin, T.~L.~Smith, T.~Karwal and M.~Kamionkowski,
Phys. Rev. Lett. \textbf{122}, no.22, 221301 (2019)
doi:10.1103/PhysRevLett.122.221301
[arXiv:1811.04083 [astro-ph.CO]].

\bibitem{Braglia:2020auw}
M.~Braglia, M.~Ballardini, F.~Finelli and K.~Koyama,
Phys. Rev. D \textbf{103}, no.4, 043528 (2021)
doi:10.1103/PhysRevD.103.043528
[arXiv:2011.12934 [astro-ph.CO]].


\bibitem{Gariazzo:2021qtg}
S.~Gariazzo, E.~Di Valentino, O.~Mena and R.~C.~Nunes,
Phys. Rev. D \textbf{106}, no.2, 023530 (2022)
doi:10.1103/PhysRevD.106.023530
[arXiv:2111.03152 [astro-ph.CO]].

\bibitem{Benisty:2024lmj}
D.~Benisty, S.~Pan, D.~Staicova, E.~Di Valentino and R.~C.~Nunes,
Astron. Astrophys. \textbf{688}, A156 (2024)
doi:10.1051/0004-6361/202449883
[arXiv:2403.00056 [astro-ph.CO]].

\bibitem{Silva:2025hxw}
E.~Silva, M.~A.~Sabogal, M.~Scherer, R.~C.~Nunes, E.~Di Valentino and S.~Kumar,
Phys. Rev. D \textbf{111}, no.12, 123511 (2025)
doi:10.1103/qqc6-76z4
[arXiv:2503.23225 [astro-ph.CO]].

\bibitem{Zhang:2025dwu}
Y.~M.~Zhang, T.~N.~Li, G.~H.~Du, S.~H.~Zhou, L.~Y.~Gao, J.~F.~Zhang and X.~Zhang,
[arXiv:2510.12627 [astro-ph.CO]].


\bibitem{DAgostino:2020dhv}
R.~D'Agostino and R.~C.~Nunes,
Phys. Rev. D \textbf{101}, no.10, 103505 (2020)
doi:10.1103/PhysRevD.101.103505
[arXiv:2002.06381 [astro-ph.CO]].

\bibitem{Adi:2020qqf}
T.~Adi and E.~D.~Kovetz,
Phys. Rev. D \textbf{103}, no.2, 023530 (2021)
doi:10.1103/PhysRevD.103.023530
[arXiv:2011.13853 [astro-ph.CO]].

\bibitem{DiValentino:2024wgi}
E.~Di Valentino, L.~Perivolaropoulos and J.~Levi Said,
doi:10.3390/universe10040184
[arXiv:2404.13981 [gr-qc]].

\bibitem{Sandoval-Orozco:2024hjm}
R.~Sandoval-Orozco, C.~Escamilla-Rivera, R.~Briffa and J.~Levi Said,
Phys. Dark Univ. \textbf{46}, 101641 (2024)
doi:10.1016/j.dark.2024.101641
[arXiv:2405.06633 [astro-ph.CO]].

\bibitem{Escamilla:2024xmz}
L.~A.~Escamilla, D.~Fiorucci, G.~Montani and E.~Di Valentino,
Phys. Dark Univ. \textbf{46}, 101652 (2024)
doi:10.1016/j.dark.2024.101652
[arXiv:2408.04354 [astro-ph.CO]].



\bibitem{Lee:2020zts}
S.~Lee,
JCAP \textbf{08}, 054 (2021)
doi:10.1088/1475-7516/2021/08/054
[arXiv:2011.09274 [astro-ph.CO]].

\bibitem{Lee:2024zcu}
S.~Lee,
Class. Quant. Grav. \textbf{42}, no.2, 025026 (2025)
doi:10.1088/1361-6382/ada2d5
[arXiv:2412.19049 [gr-qc]].

\bibitem{Lee:2025osx}
S.~Lee,
Phys. Dark Univ. \textbf{48}, 101947 (2025)
doi:10.1016/j.dark.2025.101947
[arXiv:2505.15838 [physics.gen-ph]].

\bibitem{Islam:2001}
J.~N.~Islam, \textit{An introduction to Mathematical Cosmology} (Cambridge University Press, 2001).

\bibitem{Ryder:2009}
L.~Ryder, \textit{Introduction to General Relativity} (Cambridge University Press, 2009).

\bibitem{Lee:2023rqv}
S.~Lee,
Mon. Not. Roy. Astron. Soc. \textbf{522}, no.3, 3248-3255 (2023)
doi:10.1093/mnras/stad1190
[arXiv:2301.06947 [astro-ph.CO]].

\bibitem{Lee:2023ucu}
S.~Lee,
Mon. Not. Roy. Astron. Soc. \textbf{524}, no.3, 4019-4023 (2023)
doi:10.1093/mnras/stad2084
[arXiv:2302.09735 [astro-ph.CO]].

\bibitem{Lee:2024kxa}
S.~Lee,
Phys. Dark Univ. \textbf{46}, 101703 (2024)
doi:10.1016/j.dark.2024.101703
[arXiv:2407.09532 [physics.gen-ph]].

\bibitem{Lee:2025vha}
S.~Lee,
[arXiv:2509.08840 [physics.gen-ph]].


\bibitem{Lee:2025aha}
S.~Lee,
Phys. Dark Univ. \textbf{49}, 101984 (2025)
doi:10.1016/j.dark.2025.101984


\bibitem{Lee:2021xwh}
S.~Lee,
[arXiv:2108.06043 [astro-ph.CO]].


\bibitem{Lee:2025kbn}
S.~Lee,
Mon.  Not.  R.  Astron.  Soc. \textbf{Accepted} (2025),
 \href{https://doi.org/10.1093/mnras/staf1890}{doi:10.1093/mnras/staf1890}.
[arXiv:2506.16022 [astro-ph.CO]].

\bibitem{Lee:2025ysg}
S.~Lee,
[arXiv:2506.18230 [astro-ph.CO]].













\end{thebibliography}

\end{document}